# Perceiving Motion Cues Inspired by Microsoft Kinect Sensor on Game Experiencing

Jiawei Xu[1] and Shigang Yue[2]
School of Computer Science
University of Lincoln
Brayford Pool, Lincoln,
United Kingdom
+44(0)1522837301
jxu@lincoln.ac.uk[1]
syue@lincoln.ac.uk[2]

Ruisheng Wang
Department of Geomatics
Engineering,
Schulich School of Engineering,
University of Calgary
Calgary, Canada
+ (403)210-9509
ruiswang@ucalgary.ca

Loo Chu Kiong
Advanced Robotics Lab
Department of Artificial Intelligence
Computer Science and Information
Technology, University of Malaya,
Kuala Lumpur, Malaysia
+(603)79676314
ckloo.um@um.edu.my

**Abstract**    This paper proposed a novel method to replace the traditional mouse controller by using Microsoft Kinect Sensor to realize the functional implementation on human-machine interaction. With human hand gestures and movements, Kinect Sensor could accurately recognize the participants' intention and transmit our order to desktop or laptop. In addition, the trend in current HCI market is giving the customer more freedom and experiencing feeling by involving human cognitive factors more deeply. Kinect sensor receives the motion cues continuously from the human's intention and feedback the reaction during the experiments. The comparison accuracy between the hand movement and mouse cursor demonstrates the efficiency for the proposed method. In addition, the experimental results on hit rate in the game of Fruit Ninja and Shape Touching proves the real-time ability of the proposed framework. The performance evaluation built up a promise foundation for the further applications in the field of human-machine interaction. The contribution of this work is the expansion on hand gesture perception and early formulation on Mac iPad.

## I.    INTRODUCTION

Mouse controller, no matter wire or wireless, receives user's instructions, such as click, drag or border selection in a 2-dimensional planar interface. Then the computer receives the signal and appears the feedback on the monitor or screen. Another case is the touch screen on the mobile, whether it is Apple iOS, Android or other intelligent systems, we touch the screen or click the logo via fingers. Both of them have the same feature, the user should touch the mouse or mobile then the agents can perceive the instructions. However, Microsoft Kinect sensor changed the definition of concept [1][2][3][4][5]. Microsoft Kinect SDK is developed by computer scientists or engineers to realize the preliminary human machine interaction application. They could sense our human body movement by perceiving the distance change in depth, which is actually a depth camera. Many open source code realized some basic ideas, such as skeleton recognition, human face surveillance, body tracking or even speech recognition. While in UK or North American market, the price of Kinect is relatively acceptable to customer, about 208 GBP or 330 US dollars, which is an ideal tool for further application.

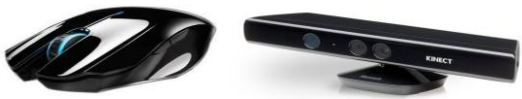

Fig.1. Replacing the tradition mouse cursor by our hand gesture provides the intuitive feeling of human-machine interaction. We designed the module by Microsoft Kinect which can capture human hand movements correctly.

## II.    RELATED WORK AND METHODOLGY

First, we analog the mouse controller in the real 3D world by sensing the human body motion, however, traditional computer mouse cursor is a two-dimensional device as it can reflect users control and instruction. A systematic review on gesture recognition or motion perception can be traced to [7], where the authors summarized and categorized previous approaches nicely. In [8], dynamic hand gestures are treated as the motion cues for visual understanding. Later on, recognition of dynamic hand gestures is realized on the conventional camera as shown in [9]. Another way, early formulation of 3D information was combined into the traditional 2D images which lead to the active appearance model in [10]. Our idea is converting the mouse into hand operation while it is free of constraints. This improves the concept on the human-machine interaction, which upgrades the mouse control by using various types of hand movements. Second, motion perception is a key point in the dynamic world. The physical motion draws our visual attention and human body responses the actions to a moving object or a dynamic agent. For example, a frisbee is flying over our head and we will try to pick it. Thus we have a corresponding reaction to the response or decision while our body moves and adjusts to a correct pose to catch it. It is a complex process and we need to finish it in a very short second. Then the second motivation for this paper is used to realize our hand movements and computer can be perceptible by this kind of motion using Microsoft Kinect sensor. In general, our work can be summarized as following: We replace the traditional mouse controller by our direct hand actions under the real 3D world, and these actions can be perceived by laptop or desktop.

## A. Depth Perception

The Kinect contains three vital pieces that work together to detect your motion and create your physical image on the screen: an RGB color VGA video camera, a depth sensor, and a multi-array microphone. The camera detects the red, green, and blue color components as well as body-type and facial features. It has a pixel resolution of 640x480 and a frame rate of 30 fps. This helps in facial recognition and body recognition. The depth sensor contains a monochrome CMOS sensor and infrared projector that help create the 3D imagery throughout the room. It also measures the distance of each point of the player's body by transmitting invisible near-infrared light and measuring its "time of flight" after it reflects off the objects [6] [11].

## B. Hand Gesture Recognition

During this process, an important factor is the motion perception. A sense of motion is crucial for the perception of our own motion in relation to other moving and static objects in the environment. While the motion is mainly composed of moving direction and motion intensity, a complete model for the motion perception can be modulated by the top-down input, which is a kind of prior knowledge, and bottom-up features, such as intensity, depth or noise[12]. For an example, while we are doing experiment on human-machine interaction, like Fruit Ninja, the fruits are coming up from the bottom line of the screen, then this prior information helps our decision and improves our estimation of the emerging objects. Another, the Kinect sensor could perceive our hand actions in the different conditions if we put ourselves under various environment, i.e., indoor or outdoor, high brightness or looming background. Table I defines our main functions of the hand gesture, each hand is responsible for particular aims.

TABLE I.    CATEGORIZATION AND MAIN FUNCTIONS OF OUR HANDS

| Left hand | Right hand |
| --- | --- |
| Click/Confirmation<br>Cutting<br>Moving | Drag/Localization<br>Keeping balance<br>Rotation |

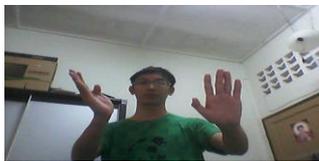

Fig. 2.  Snapshot of the hands gesture during the experiments

A framework illustrated the flowchart of our coding module is shown in figure 2. Figure 3 indicate our flowchart scheme in general. While Table II reflects the basic order by our hand gestures, the reference data is the traditional mouse cursor.

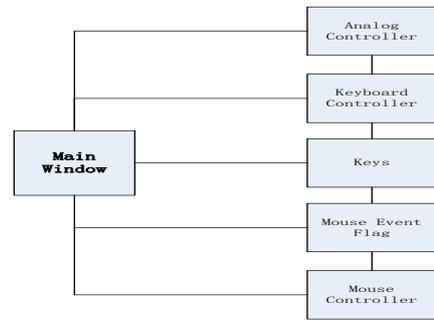

Fig. 3.  The schematic module is composed by 5 sections, the main window aggregates each specific functions.

TABLE II.    ACCURACY PERCENTAGE BY THE HAND MOVEMENTS COMPARED WITH THE TRADITIONAL MOUSE (GROUND-TRUTH)

| | Click | Cutting | Drag | Balancing | Rotation |
| --- | --- | --- | --- | --- | --- |
| A | 98.75 | 78.43 | 81.62 | 83.34 | 85.24 |
| B | 84.27 | 86.37 | 88.36 | 84.41 | 72.42 |
| C | 80.70 | 82.40 | 84.99 | 83.88 | 78.83 |
| D | 99.27 | 97.28 | 97.29 | 94.47 | 83.67 |
| E | 99.18 | 97.78 | 94.57 | 84.84 | 71.13 |
| F | 99.27 | 98.75 | 97.96 | 94.77 | 87.23 |
| G | 99.18 | 97.94 | 96.26 | 84.67 | 68.27 |
| H | 99.27 | 97.75 | 96.46 | 93.86 | 86.38 |
| I | 99.18 | 97.46 | 94.24 | 87.23 | 73.92 |
| J | 99.27 | 77.12 | 97.96 | 94.56 | 87.23 |
| K | 99.21 | 98.32 | 97.71 | 94.03 | 82.32 |
| L | 99.29 | 98.00 | 96.56 | 91.05 | 80.02 |
| M | 98.71 | 97.92 | 97.92 | 93.72 | 82.33 |
| N | 80.84 | 79.91 | 78.44 | 70.33 | 68.51 |

## III.    EXPERIMENTS

Our experiment environment is described as below. CPU is Intel Core(TM) i5 D3 480M Dual Core Processor, 4G RAM, GPU is NVIDIA GeForce GT445 1G with a 15.6" Monitor. Visual Studio 2010 is mounted on the laptop and Microsoft Kinect Sensor has been successfully set up.

14 participants (from A to N) joined our experiments for testing the accuracy comparing with reference data, using the traditional mouse. Due to the page limitation for this conference, here we only demonstrated our core code below showing the function realization in Table III.

TABLE III.    CORE CODE ON MOUSE CONTROLLER

```
using System;
using System.Collections.Generic;
using System.Linq;
using System.Text;
using System.Threading.Tasks;
using System.Windows;
using System.Runtime.InteropServices;
namespace AnalogController
{
        public class MouseContorller
    {
        /// <summary>
```

```
/// Mimic Mouse Control
/// </summary>
/// <param name="dwFlags">Mimic mouse data</param>
/// <param name="dx">x</param>
/// <param name="dy">y</param>
/// <param name="dwData">0</param>
/// <param name="dwExtraInfo">0</param>
[DllImport("user32.dll", EntryPoint = "mouse_event")]
public static extern void mouse_event(MouseEventFlag dwFlags, int dx,
int dy, int dwData, int dwExtraInfo);
/// <summary>
/// Setting Mouse Position
/// </summary>
/// <param name="X">X</param>
/// <param name="Y">y</param>
/// <returns></returns>
[DllImport("user32.dll")]
public static extern bool SetCursorPos(int X, int Y);
/// <summary>
/// Obtain the mouse position
/// </summary>
/// <param name="pt"></param>
/// <returns></returns>
[DllImport("user32.dll")]
public static extern bool GetCursorPos(out Point pt);
public double MoveWidth { get; set; }
public double MoveHeight { get; set; }
private double screenWidth;
private double screenHeight;
/// <summary>
/// Setting the width and height of our hand movement
/// </summary>
/// <param name="moveWidth">width by meter</param>
/// <param name="moveHeight">height by meter</param>
public MouseContorller(double moveWidth, double moveHeight)
{
    this.MoveWidth = moveWidth;
    this.MoveHeight = moveHeight;
    screenWidth = SystemParameters.PrimaryScreenWidth;
    screenHeight = SystemParameters.PrimaryScreenHeight;
}
/// <summary>
/// Setting the mouse position by the relative position on the screen , the
reference point is the point of the upper left corner
/// </summary>
/// <param name="OriginPointX">reference point X</param>
/// <param name="OriginPointY">reference point Y</param>
/// <param name="TargetPointX">target point X</param>
/// <param name="TargetPointY">target point Y</param>
public void SetCursorPosBySkeletonPoint(double OriginPointX,double
OriginPointY,double TargetPointX,double TargetPointY)
{
    int mouseX, mouseY;
    double dWidth = TargetPointX - OriginPointX;
    if (dWidth > MoveWidth) mouseX = 65536;
    else if (dWidth < 0) mouseX = 0;
    else mouseX = Convert.ToInt32(dWidth / MoveWidth * 65536);
    double dHeight = OriginPointY - TargetPointY;
    if (dHeight > MoveHeight) mouseY = 65536;
    else if (dHeight < 0) mouseY = 0;
    else mouseY = Convert.ToInt32(dHeight / MoveHeight * 65536);
    mouse_event(MouseEventFlag.Absolute | MouseEventFlag.Move,
mouseX, mouseY, 0, 0);
}
    /// </summary>
/// <param name="IsPress"> </param>
public void PressLeftkey(bool IsPress)
{
    if (IsPress) mouse_event(MouseEventFlag.LeftDown, 0, 0, 0, 0);
    else mouse_event(MouseEventFlag.LeftUp, 0, 0, 0, 0);
}
}
```

```
}
```

The code performs relatively robustly and we have tested it on Fruit Ninja and Shape games. The results are illustrated as below. The precision is an important factor to measure our work. After testing on these games, we find the localization and sensitivity are acceptable, the hit rate is almost same comparing with the traditional mouse controller, which are 78 and 80, respectively. (Using Kinect Sensor/ Using mouse controller). While a detailed performance evaluation will be discussed in section IV. Table IV and Table V gives us the quantitative analysis.

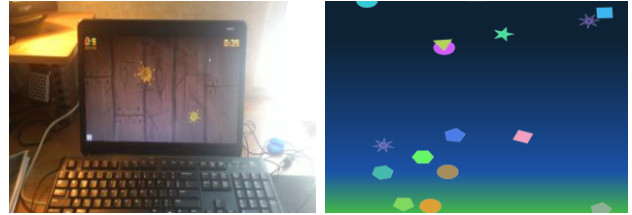

Fig. 4. Our testing is implemented on Fruit Ninja and Shape Game. We stand before the laptop about 5 meters away and replace the traditional mouse controller by our hands gestures.

## IV. DISCUSSION

The primary goal of the current research is to explore the further application on the human-machine interaction. We notice that, a similar technology has already been built and released recently by Leap Motion, Inc (*https://www.leapmotion.com/product*). However, there is a difference in that Leap Motion created his own hardware, while our project is an adaptation of existing hardware. Hence there is the novelty and potential for this interesting research and technology. We further compare our method with Android mobile application by taking average scores over 100 games of Fruit Ninja and Shape Game.

TABLE IV.     Average scoring over 100 games of Fruit Ninja, the middle column indicates the module developed by our method; right column is the score on Mac iPad by 14 voluntary participants.

| Participants | Average Score by Hand Gestures | Average Score Testing on iPad |
|---|---|---|
| A | 178 | 159 |
| B | 185 | 167 |
| C | 169 | 178 |
| D | 192 | 179 |
| E | 178 | 191 |
| F | 193 | 175 |
| G | 191 | 194 |
| H | 197 | 175 |
| I | 187 | 196 |
| J | 179 | 172 |
| K | 191 | 198 |
| L | 179 | 188 |
| M | 182 | 189 |
| N | 180 | 179 |

To summarize, the performance of our module outperforms the average score testing on iPad. The experiment results positively reflect the superiority hand gesture with respect to Fruit Ninja. We also compared our method on Kinect Shape Game by using our module and iPad, experiment results also promise a higher performance. Here we illustrated the hit rate in table V.

TABLE V.     AVERAGE HIT RATE OVER 100 GAMES OF SHAPE GAME , THE MIDDLE COLUMN INDICATES THE MODULE DEVELOPED BY OUR METHOD; RIGHT COLUMN IS THE SCORE ON MAC IPAD BY 14 VOLUNTARY PARTICIPANTS.

| Participants | Average Hit Rate Testing on Desktop | Average Hit Rate Testing on Mobile |
|---|---|---|
| A | 76% | 79% |
| B | 68% | 65% |
| C | 74% | 72% |
| D | 85% | 81% |
| F | 91% | 95% |
| G | 72% | 84% |
| H | 87% | 77% |
| I | 75% | 76% |
| J | 82% | 85% |
| K | 67% | 65% |
| L | 74% | 73% |
| M | 68% | 69% |
| N | 78% | 75% |

Furthermore, we compared our method with other state-of-the-art methods to evaluate this method's advantage and disadvantage. Compared with [13], the author advised a method based on the ZCam and an SVM-SMO classifier. Furthermore, the authors in [14] proposed the hand gesture using a range camera with a satisfactory real-time ability. In [15], motion tracking using a range camera and the mean-shift algorithm was combined together to capture the hand gestures. There are still a few methods but hereby we compared these 3 approaches as we simulated them thoroughly on 14 subjects by giving the same game experiencing. Here below is our result indicated in table VI.

TABLE VI.     COMPARISONS WITH OTHER STATE-OF-THE-ART METHODS.

| Participants | Average Recognition Accuracy | Miss Rate |
|---|---|---|
| [13] | 0.77 | 0.19 |
| [14] | 0.74 | 0.21 |
| [15] | 0.79 | 0.17 |
| Proposed method | 0.85 | 0.14 |

## V.  CONCLUSION

The idea of our experiments is straightforward and successful by using Microsoft Kinect SDK skillfully. Actually, application development can be more sophisticated than this and we are still on the primary stage of human-machine interaction. There exist more high level and more cognitive processes in future [16] [17]. We have also tried other applications in the lab, human face tracking, and online fitting room by Unity 3D or the latest innovation, Leap Motion. Anyway, intelligent human-machine interaction by using advanced equipment is the state-of-the-art trend in the upcoming years [18]. Another, as motion cues are the key factor dominated our visual perception by our previous research, it is essential for us to develop new modules by integrating motion cues [19].

Future work will be on the combination of motion cues which can be extracted quickly and robustly to identify attentive objects that are most relevant to the entity that possess the vision system [20] [21] [22].